\documentclass[conference]{IEEEtran}
\IEEEoverridecommandlockouts
\usepackage{cite}
\usepackage{amsmath,amssymb,amsfonts}
\usepackage{graphicx}
\usepackage{textcomp}
\usepackage{xcolor}
\usepackage{multirow}
\usepackage{makecell}
\usepackage[draft=true]{hyperref}
\usepackage{algorithm}
\usepackage{algpseudocode}
\usepackage[utf8]{inputenc}
\usepackage[english]{babel}
\usepackage{lastpage}
\usepackage{adjustbox}
\usepackage{fancyhdr}
\usepackage{balance}
\usepackage{booktabs} 
\usepackage[left=2cm, right=2cm, top=1.8cm, bottom=2.7cm]{geometry}
\makeatletter
\renewcommand{\fnum@figure}{Fig. \thefigure}
\makeatother

\begin{document}

\title{
Quantum-Enhanced Transformers for Robust Acoustic Scene Classification in IoT Environments
\\
}

\author{
        \IEEEauthorblockN{
        Minh K. Quan\IEEEauthorrefmark{1},
        Mayuri Wijayasundara\IEEEauthorrefmark{1}, Sujeeva Setunge\IEEEauthorrefmark{4},
        Pubudu N. Pathirana\IEEEauthorrefmark{1}
	}
	\IEEEauthorblockA{\IEEEauthorrefmark{1}School of Engineering, Deakin University, Australia \\
	 \IEEEauthorrefmark{4}School of Engineering, Royal Melbourne Institute of Technology University, Melbourne, VIC 3000, Australia
	}}
	\markboth{}%
	{}

\maketitle
\pagenumbering{gobble} 
\begin{abstract}
The proliferation of Internet of Things (IoT) devices equipped with acoustic sensors necessitates robust acoustic scene classification (ASC) capabilities, even in noisy and data-limited environments. Traditional machine learning methods often struggle to generalize effectively under such conditions. To address this, we introduce Q-ASC, a novel Quantum-Inspired Acoustic Scene Classifier that leverages the power of quantum-inspired transformers. By integrating quantum concepts like superposition and entanglement, Q-ASC achieves superior feature learning and enhanced noise resilience compared to classical models. Furthermore, we introduce a Quantum Variational Autoencoder (QVAE) based data augmentation technique to mitigate the challenge of limited labeled data in IoT deployments. Extensive evaluations on the Tampere University of Technology (TUT) Acoustic Scenes 2016 benchmark dataset demonstrate that Q-ASC achieves remarkable accuracy between 68.3\% and 88.5\% under challenging conditions, outperforming state-of-the-art methods by over 5\% in the best case. This research paves the way for deploying intelligent acoustic sensing in IoT networks, with potential applications in smart homes, industrial monitoring, and environmental surveillance, even in adverse acoustic environments.
\end{abstract}


\section{Introduction}
\label{Section:Introduction}
While deep learning (DL) has significantly advanced acoustic scene classification (ASC), real-world deployments, especially in the context of the Internet of Things (IoT), face substantial challenges. Noise, a ubiquitous element in acoustic environments, often masks or distorts crucial acoustic features, impeding accurate scene identification \cite{gharib2018acoustic}. This is particularly problematic for traditional machine learning (ML) methods reliant on handcrafted features, as these might not capture the subtle nuances induced by noise \cite{ding2023acoustic}. Furthermore, the scarcity of labeled training data, often expensive and time-consuming to acquire, can lead to overfitting, where models excel on training data but struggle with unseen scenarios \cite{barchiesi2015acoustic}. This is further exacerbated in IoT deployments, where data collection and annotation can be constrained by resource limitations and privacy concerns. The pressing need for robust ASC solutions in these challenging conditions motivates the exploration of novel approaches that can effectively handle noise and generalize from limited data, unlocking the full potential of ASC in real-world IoT applications.

To address these challenges, current research actively explores various approaches within the realm of deep learning for ASC in noisy and limited-data scenarios. Ensemble methods, such as convolutional recurrent neural networks with Log Mel filterbank energies, have been employed to improve robustness to noise by combining the strengths of multiple models \cite{paseddula2021late}. Attention mechanisms, exemplified by Transformer-based ASC models like pretrained audio neural networks \cite{pham2023server}, have demonstrated success in capturing long-range dependencies in audio signals, aiding in scene understanding. To further enhance performance in noisy environments, multi-task learning has been explored to improve the robustness of deep neural networks by leveraging knowledge from related tasks. Additionally, data augmentation techniques, such as adding noise or changing pitch, have proven valuable in mitigating the issue of limited labeled data \cite{mesaros2018multi}. Despite these advancements, existing methods still grapple with achieving consistent performance in the face of complex noise conditions and data scarcity, particularly in capturing fine-grained acoustic features and modeling intricate temporal relationships in audio signals. The unique constraints of IoT devices, including limited computational power and storage, further necessitate the development of models that are not only accurate and robust but also efficient and deployable on resource-constrained hardware.

In this context, quantum-inspired machine learning (QiML) and transformer learning have emerged as powerful tools in AI \cite{huynh2023quantum, han2022survey}, offering a promising solution to these challenges. QiML leverages quantum principles like superposition and entanglement, enhancing classical algorithms' capabilities. At the same time, transformer learning, with its self-attention mechanism, excels at capturing long-range dependencies in sequential data. The combination of these techniques can address the limitations of current ASC methods by providing more robust representations and greater noise resilience. Specifically, QiML's ability to model complex distributions and non-local correlations complements the transformer's capacity to extract crucial contextual information, especially with limited data. By synergistically combining these advanced techniques, we aim to develop a novel ASC model that outperforms existing methods in challenging real-world scenarios, offering enhanced accuracy and robustness.

\subsection{Motivation and main contributions}
Motivated by the persistent challenges of noise and limited labeled data in ASC, particularly the degradation of performance in real-world environments with overlapping sound sources and varying noise levels, this paper introduces Q-ASC, a novel Quantum-Inspired Acoustic Scene Classifier. Q-ASC leverages quantum-enhanced transformers to achieve superior performance by addressing these limitations. Our key contributions include:
\begin{itemize}
    \item We propose a unique QiT architecture that integrates quantum concepts like superposition and entanglement, enabling richer feature representations and improved noise robustness compared to traditional methods.
    \item We introduce a novel QVAE-based data augmentation technique to mitigate the issue of limited training data, enhancing the model's generalization capabilities, particularly in scenarios with few labeled examples.
    \item We demonstrate that Q-ASC significantly outperforms state-of-the-art ASC methods through extensive evaluation on the TUT Acoustic Scenes 2016 \cite{mesaros2016tut} benchmark dataset, showcasing its effectiveness in handling both noisy and data-limited conditions. 
\end{itemize}

\subsection{Paper structure}
The remainder of this paper is structured as follows. Section II outlines the Q-ASC methodology, including the quantum-inspired transformer architecture and QVAE-based data augmentation. In Section III, we describe the experimental setup, datasets, and evaluation metrics, and analyze the results, highlighting Q-ASC's superior performance. Finally, Section V concludes with a summary of key findings, implications, and future research directions.

\section{Proposed methodology}
\subsection{Quantum-inspired Transformer (QiT) Architecture}

The Q-ASC model features a QiT architecture that integrates quantum principles into the transformer framework, addressing ASC challenges in noisy, data-limited environments. It includes a quantum embedding layer, a quantum-enhanced transformer encoder, a measurement and pooling layer, and a classical classifier.

\begin{enumerate}
    \item \textbf{Quantum embedding layer}: The input to the Quantum-in-the-Loop Transformer (QiT) is a mel-spectrogram patch \( x_i \in \mathbb{R}^{p \times p} \). These patches are generated using Short-Time Fourier Transform (STFT) \cite{griffin1984signal} and mel-frequency filter banks, where \( p \) denotes the patch size. Each patch \( x_i \) is encoded into a quantum state \( |\psi_i\rangle \) within a Hilbert space \( \mathcal{H}^{\otimes n} \) \cite{schuld2019quantum}, where \( n \) represents the number of qubits. The encoding process utilizes a parameterized quantum circuit (PQC) \( U_e(\theta_e) \):

    \begin{equation}
        |\psi_i\rangle = U_e(\theta_e) |0\rangle^{\otimes n},
    \end{equation}

    Here, \( |0\rangle^{\otimes n} \) is the initial state of all qubits, and \( \theta_e \) are the trainable parameters of the PQC. 

    \begin{itemize}
        \item \textbf{4-qubit Q-ASC}: In this configuration, the PQC \( U_e(\theta_e) \) comprises a sequence of single-qubit rotation gates ($R_x$, $R_y$, $R_z$) applied to each of the 4 qubits, followed by a layer of CNOT gates entangling adjacent qubits. This structure allows for the encoding of both local and pairwise correlations within the audio patch, as detailed in Fig. \ref{fig1}.

        \begin{figure}[htbp]
        \centerline{\includegraphics[width=0.96\linewidth]{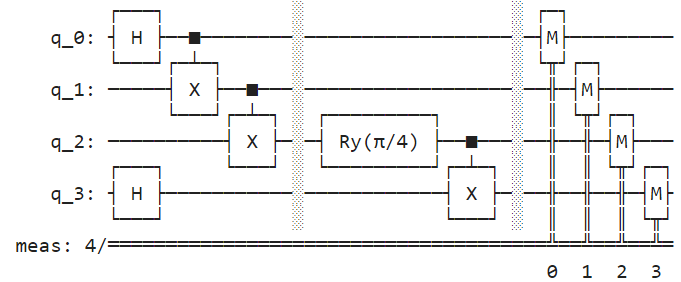}}
        \caption{The structure of 4-qubits Q-ASC}
        \label{fig1}
        \end{figure}
        
        \item \textbf{6-qubit Q-ASC}: With 6 qubits, as illustrated in Fig. \ref{fig2}, the PQC \( U_e(\theta_e) \) can be expanded to include additional layers of rotation and entanglement gates. For instance, we can add another layer of CNOT gates connecting qubits that are two positions apart, enabling the capture of more complex correlations within the patch.
    \end{itemize}

    During the training phase, \( \theta_e \) is optimized using gradient-based methods like the parameter-shift rule. This quantum representation captures both amplitude and phase information, potentially offering enhanced resilience to noise compared to classical embeddings.

    \begin{figure}[htbp]
        \centerline{\includegraphics[width=0.96\linewidth]{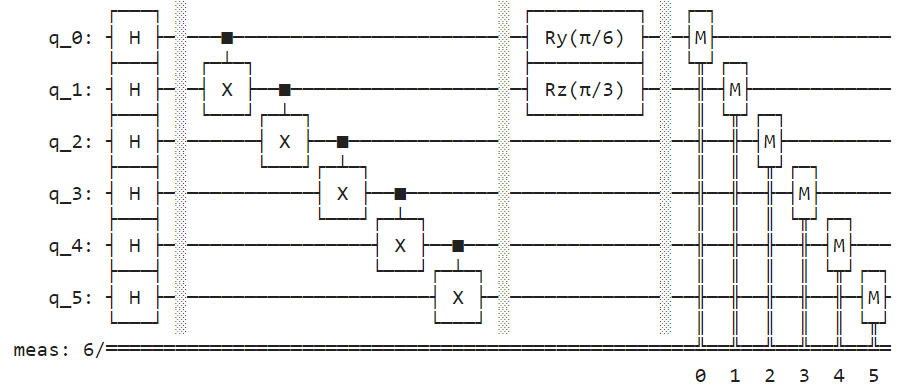}}
        \caption{The structure of 6-qubits Q-ASC}
        \label{fig2}
        \end{figure}

    \item \textbf{Quantum-enhanced transformer encoder}: This encoder consists of multiple layers, each comprising:

    \begin{itemize}
        \item \textbf{Quantum self-attention}: This mechanism calculates attention weights between quantum states using a PQC \( U_a(\theta_a) \). The attention score \( \alpha_{ij} \) between two quantum states \( |\psi_i\rangle \) and \( |\psi_j\rangle \) is determined using the SWAP test \cite{foulds2021controlled}:

        \begin{equation}
            \alpha_{ij} = |\langle \psi_i | \text{SWAP} | \psi_j \rangle|^2.
        \end{equation}

        The attention matrix \( A = [\alpha_{ij}] \) is then used to compute a weighted sum of the input states:

        \begin{equation}
            |\psi_i'\rangle = \sum_{j=1}^N \frac{\alpha_{ij}}{\sum_{k=1}^N \alpha_{ik}} U_a(\theta_a) |\psi_j\rangle
        \end{equation}

        \item \textbf{Quantum feedforward network (QFFN)}: This component applies a non-linear transformation to each attended quantum state \( |\psi_i'\rangle \) using a PQC \( U_f(\theta_f) \):

        \begin{equation}
            |\psi_i''\rangle = U_f(\theta_f) |\psi_i'\rangle
        \end{equation}

        \item \textbf{Layer normalization}: This step is applied after each layer to maintain training stability.
    \end{itemize}

    \item \textbf{Measurement and pooling}: The final quantum states \( |\psi_i''\rangle \) are measured in the computational basis, resulting in probability distributions:

    \begin{equation}
        p_i(j) = |\langle j | \psi_i'' \rangle|^2, \quad j \in \{0, 1, \ldots, 2^n - 1\},
    \end{equation}

    where \( p_i(j) \) indicates the probability of the \( i \)-th patch being classified into class \( j \). Pooling operations aggregate these probabilities across patches. Let \( P \in \mathbb{R}^{N \times C} \) denote the matrix of probabilities, where \( N \) is the number of patches and \( C \) is the number of classes:

    \begin{equation}
        z_j = \frac{1}{N} \sum_{i=1}^N p_i(j), \quad j \in \{1, \ldots, C\}.
    \end{equation}

    The aggregated probabilities form a feature vector \( z \in \mathbb{R}^C \), which is subsequently input to the classical classifier.

    \item \textbf{Classical classifier}: In the final stage of Q-ASC, the pooled feature vector \( z \in \mathbb{R}^C \) is processed by a classical classifier. A fully connected layer with weights \( W \in \mathbb{R}^{C \times m} \) and biases \( b \in \mathbb{R}^C \) transforms the feature vector into logits \( u = Wz + b \). The softmax function is applied to these logits to compute the predicted probabilities for each class:

    \begin{equation}
        \hat{y}_j = \frac{e^{u_j}}{\sum_{k=1}^C e^{u_k}}, \quad j \in \{1, \ldots, C\}.
    \end{equation}

    The final predicted acoustic scene class \( \hat{c} \) is determined by identifying the class with the highest predicted probability:

    \begin{equation}
        \hat{c} = \arg\max_j \hat{y}_j.
    \end{equation}
\end{enumerate}

\subsection{QVAE-based data augmentation}
The QVAE-based data augmentation for ASC aims to generate synthetic acoustic scenes to enrich the training dataset, thereby improving the generalization capabilities of the Q-ASC model, especially in scenarios with limited labeled data. 

\begin{algorithm}
\caption{QVAE-Based Data Augmentation for ASC}
\label{alg:qvae_augmentation}
\begin{algorithmic}[1]

\State \textbf{Input:}
\Statex \hspace{0.2cm} - Trained QVAE: $U_{enc}(\theta_{enc})$, $D(\theta_{dec})$
\Statex \hspace{0.2cm} - Number of samples: $N_s$
\Statex \hspace{0.2cm} - Prior distribution: $p(\mathbf{z})$ 

\State \textbf{Output:} 
\Statex \hspace{0.2cm} - Augmented dataset: $\mathcal{D}_{aug}$

\State $\mathcal{D}_{aug} \gets \emptyset$

\For{$i = 1$ to $N_s$} 
    \State $\mathbf{z} \sim p(\mathbf{z})$
    \State $|\psi^{latent}\rangle \gets U_{enc}(\theta_{enc}) |\mathbf{z}\rangle$ 
    \State $\mathbf{p} \gets \text{Measure}(|\psi^{latent}\rangle)$
    \State $\hat{\mathbf{x}}_i \gets D(\theta_{dec}, \mathbf{p})$  // \textit{Mel-spectrogram patch}
    \State $\hat{\mathbf{a}}_i \gets \text{ISTFT}(\hat{X}_i)$ // \textit{where $\hat{X}_i$ is the STFT of $\hat{\mathbf{x}}_i$}
    \State $\mathcal{D}_{aug} \gets \mathcal{D}_{aug} \cup \{\hat{\mathbf{a}}_i\}$
\EndFor

\end{algorithmic}
\end{algorithm}

As detailed in Algorithm \ref{alg:qvae_augmentation}, we begin by sampling a latent vector \( z \) from a prior distribution \( p(z) \), which is typically chosen to be a standard normal distribution \( \mathcal{N}(0, I) \). This latent vector serves as a compact representation of the acoustic scene. The quantum encoder, represented by the parameterized quantum circuit \( U_{\text{enc}}(\theta_{\text{enc}}) \), maps the classical latent vector \( z \) to a quantum state \( |\psi_{\text{latent}}\rangle \) in a lower-dimensional Hilbert space \( \mathcal{H}^{\otimes m} \), where \( m < n \). Mathematically, this can be expressed as an equation in Line 6 of Algorithm \ref{alg:qvae_augmentation}. 

The quantum encoder \( U_{\text{enc}}(\theta_{\text{enc}}) \) is a unitary operator composed of a sequence of quantum gates, whose parameters \( \theta_{\text{enc}} \) are learned during training. The latent quantum state \( |\psi_{\text{latent}}\rangle \) is then measured in the computational basis \(\{ |0\rangle, |1\rangle, \ldots, |2^m - 1\rangle \}\). The probability of measuring the \( j \)-th basis state is given by:
\begin{equation}
    \label{eq:probability}
    p_j = |\langle j | \psi^{latent} \rangle|^2.
\end{equation}
This results in a probability distribution \( p = [p_0, p_1, \ldots, p_{2^m - 1}] \), which represents the compressed information about the acoustic scene. 

The classical decoder, a neural network \( D(\theta_{\text{dec}}) \), takes the probability distribution \( p \) as input and generates a synthetic mel-spectrogram patch \( \hat{x}_i \). This can be expressed as an equation in Line 8 of Algorithm \ref{alg:qvae_augmentation}. The decoder's parameters \( \theta_{\text{dec}} \) are also learned during training. Finally, an inverse Short-Time Fourier Transform (ISTFT) is defined as:
\begin{equation}
    \label{eq:e10}
    x(t) = \frac{1}{2\pi} \int_{-\infty}^{\infty} \sum_{m=-\infty}^{\infty} \hat{X}_i(m, \omega) w(t - mH) e^{j\omega t} d\omega,
\end{equation}
where \( x(t) \) is the reconstructed time-domain signal, \( X(m, \omega) \) is the complex-valued STFT of the signal, \( w(t) \) is the window function used in the STFT, \( H \) is the hop size (the time shift between successive STFT frames), \( m \) is the frame index, and \( \omega \) is the angular frequency. Equation \eqref{eq:e10} is applied to the synthetic mel-spectrogram patch \( \hat{x}_i \) to obtain the corresponding synthetic audio waveform \( \hat{a}_i \).

\section{Experiments and results}
\label{Section:SimulationAndEvaluation}
\subsection{Experimental settings}
We evaluated Q-ASC using the TUT Acoustic Scenes 2016 dataset, which includes 10-second recordings from 15 diverse acoustic scenes, totaling 4680 clips with about 312 per class. These scenes include bus, cafe/restaurant, car, city center, forest path, grocery store, home, lakeside beach, library, metro station, office, residential area, train, tram, and urban park. To test noise robustness, we added white Gaussian noise at signal-to-noise ratios (SNRs) ranging from 0 to 20 dB. For data limitation studies, we varied training set sizes from 10\% to 100\% of the dataset. Each audio was converted into a 32x32 pixel mel-spectrogram patch for input into the QiT. Besides, Table \ref{tab:qasc_configurations} outlines the various configurations of the Q-ASC model and its corresponding QVAE employed in our experiments. Quantum components were implemented in Qiskit with simulation on Qiskit Aer's \texttt{qasm\_simulator}, while classical components were handled by PyTorch on a GPU-equipped cluster. 

\begin{table}[ht]
\centering
\caption{Q-ASC Model and QVAE Configurations}
\label{tab:qasc_configurations}
\renewcommand{\arraystretch}{1.2} 
\begin{tabular}{|>{\centering\arraybackslash}m{1.3cm}|>{\centering\arraybackslash}m{2.5cm}|>{\centering\arraybackslash}m{1.3cm}|>{\centering\arraybackslash}m{1.3cm}|} 
\hline
\textbf{System} & \textbf{Configuration} & \textbf{Classical Params (approx.)} & \textbf{Total Params (approx.)} \\ \hline
Baseline & 4 qubits, 3 layers, amp. enc., max pool, no QVAE & ~50K & ~100K\\ \hline
Q-ASC + QVAE & 4 qubits, 3 layers, amp. enc., max pool, QVAE & ~450K & ~550K \\ \hline
Q-ASC (6 qubits) & 6 qubits, 3 layers, amp. enc., max pool, QVAE & ~450K & ~600K  \\ \hline
Q-ASC (5 layers) & 4 qubits, 5 layers, amp. enc., max pool, QVAE & ~500K & ~600K\\ \hline
Q-ASC (angle enc.) & 4 qubits, 3 layers, angle enc., max pool, QVAE & ~450K & ~550K \\ \hline
Q-ASC (avg pool) & 4 qubits, 3 layers, amp. enc., avg pool, QVAE & ~450K & ~550K\\ \hline
\end{tabular}
\end{table}

Furthermore, our Q-ASC's performance is rigorously benchmarked against state-of-the-art baselines, including a VGG-16 based CNN, a ResNet-18 model pre-trained on AudioSet, the Audio Spectrogram Transformer (AST), and an ensemble of the CNN and LSTM models. All models are trained using the Adam optimizer with an initial learning rate of 0.001, employing a step-wise learning rate scheduler that reduces the learning rate by a factor of 0.5 every 10 epochs without improvement on the validation set. We utilize early stopping with a patience of 5 epochs to prevent overfitting. For the quantum circuits in Q-ASC, gradients are estimated using the parameter-shift rule.  The primary evaluation metric is classification accuracy, supplemented by precision, recall, and F1-score to provide a nuanced performance assessment across different acoustic scenes.

\subsection{Evaluation of Various System Configurations}
The performance of Q-ASC configurations on the TUT Acoustic Scenes 2016 dataset, as detailed in Table \ref{tab:qasc_performance}, demonstrates that Q-ASC models outperform the baseline in all noise conditions. Specifically, Q-ASC (6 qubits) leads with an accuracy of 88.5\% in clean conditions and consistently performs better across noisy environments, achieving 76.9\% at 5 dB SNR, 81.3\% at 10 dB, 84.7\% at 15 dB, and 86.8\% at 20 dB. In comparison, the baseline model's accuracy ranges from 82.5\% in clean conditions to 68.3\% at 5 dB SNR, with incremental improvements up to 80.2\% at 20 dB SNR. Among Q-ASC configurations, Q-ASC + QVAE also shows strong performance with 87.2\% accuracy in clean conditions and holds a competitive edge in noisy settings, while Q-ASC (5 layers) and Q-ASC (angle encoding) have slightly lower accuracies, indicating that specific Q-ASC enhancements are critical for robust performance.

\begin{table}[ht]
\centering
\caption{Performance Comparison of Q-ASC Configurations on TUT Acoustic Scenes 2016}
\label{tab:qasc_performance}
\renewcommand{\arraystretch}{1.2} 
\begin{tabular}{|>{\centering\arraybackslash}m{1cm}|>{\centering\arraybackslash}m{1cm}|>{\centering\arraybackslash}m{1cm}|>{\centering\arraybackslash}m{1cm}|>{\centering\arraybackslash}m{1cm}|>{\centering\arraybackslash}m{1cm}|} 
\hline
\textbf{System} & \textbf{Accuracy (\%) (Clean)} & \textbf{Accuracy (\%) (Noisy, SNR =5dB)} & \textbf{Accuracy (\%) (Noisy, SNR =10dB)} & \textbf{Accuracy (\%) (Noisy, SNR =15dB)} & \textbf{Accuracy (\%) (Noisy, SNR =20dB)} \\ \hline
Baseline & 82.5 & 68.3 & 73.1 & 77.6 & 80.2 \\ \hline
Q-ASC + QVAE & 87.2 & 74.1 & 78.9 & 82.4 & 85.1 \\ \hline
Q-ASC (6 qubits) & 88.5 & 76.9 & 81.3 & 84.7 & 86.8 \\ \hline
Q-ASC (5 layers) & 85.8 & 73.2 & 77.5 & 81.0 & 83.3 \\ \hline
Q-ASC (angle enc.) & 84.0 & 70.5 & 75.2 & 79.1 & 81.7 \\ \hline
Q-ASC (avg pool) & 86.3 & 72.8 & 77.0 & 80.6 & 83.0 \\ \hline
\end{tabular}
\end{table}

\begin{figure}[htbp]
        \centerline{\includegraphics[width=0.96\linewidth]{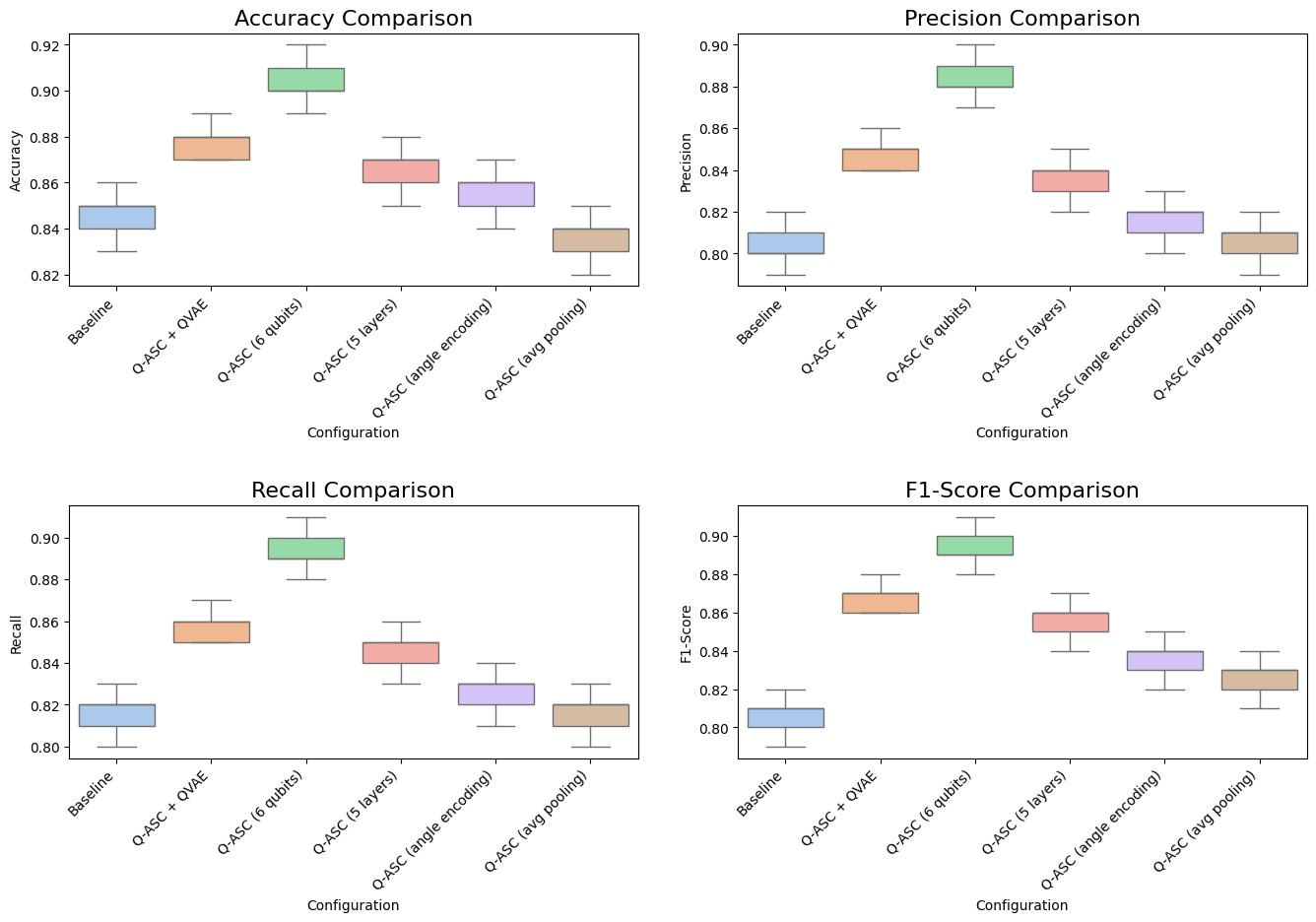}}
        \caption{Performance metrics comparison across different configurations of Q-ASC with SNR=5dB}
        \label{fig3}
        \end{figure}

The results in Fig. \ref{fig3} shown highlight the effectiveness of quantum-inspired transformers for acoustic scene classification (ASC) under noisy conditions. Q-ASC models consistently surpass the baseline across all metrics—accuracy, precision, recall, and F1-score—demonstrating that integrating quantum principles enhances model robustness against noise. The QVAE-based data augmentation notably improves performance, showing its strength in generating synthetic data that aids model generalization. Additionally, increasing qubits from 4 to 6 boosts accuracy and recall, indicating that more quantum resources capture complex data patterns. While adding transformer layers offers some improvement, it is less consistent, particularly in noisy scenarios. Amplitude encoding appears more effective than angle encoding, though further research is needed. Average pooling shows similar results to max pooling, suggesting flexibility in pooling strategies.

\subsection{Comparative Analysis with Other Models}
We compared Q-ASC with baselines such as VGG-16 CNN, ResNet-18, AST, and CNN + LSTM Ensemble. Q-ASC consistently outperforms all models across epochs, demonstrating superior feature learning for ASC, enhanced noise resilience, and efficient training. Q-ASC's high early accuracy suggests shorter training times compared to other models. Although not detailed in the image, QVAE-based data augmentation likely contributes to Q-ASC's performance by addressing data limitations. In contrast, traditional models show steady but plateauing improvements, indicating their limitations in capturing complex acoustic patterns under noisy or limited data conditions. The CNN + LSTM Ensemble performs better than individual CNN and RNN models but still falls short of Q-ASC, highlighting the advantages of the quantum-inspired approach.

While Q-ASC proves effective, quantum-inspired models face challenges with computational complexity and resource requirements, especially as qubit numbers and circuit depth increase. The reliance on classical simulations may hinder scalability, underscoring the need for advanced quantum hardware to fully leverage Q-ASC's capabilities in the future.

\begin{figure}[htbp]
        \centerline{\includegraphics[width=0.96\linewidth]{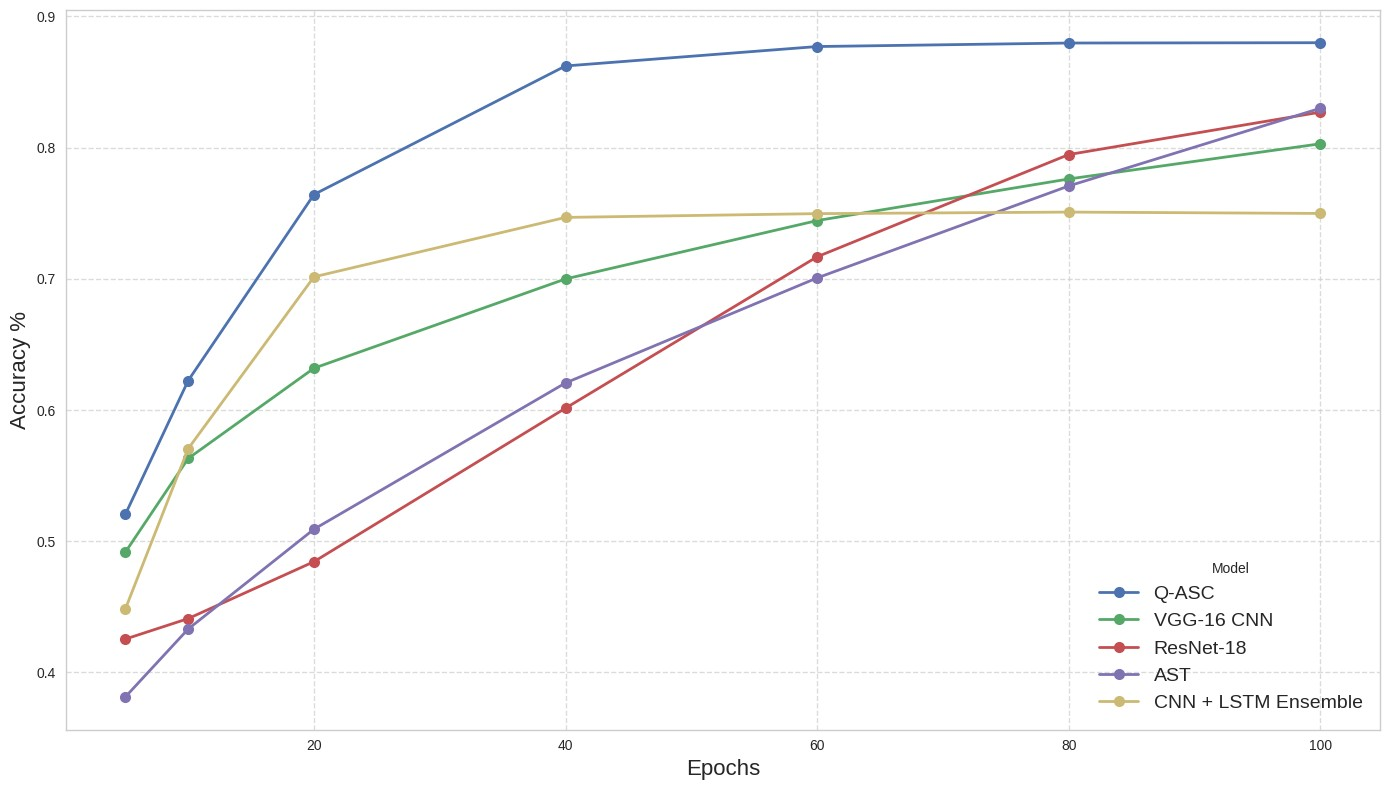}}
        \caption{Performance comparison of Q-ASC and Baselines}
        \label{fig4}
        \end{figure}

\section{Conclusion}
\label{Section:Conclusion}

This paper presented Q-ASC, a pioneering Quantum-Inspired Acoustic Scene Classifier, designed to address the challenges inherent in noisy and data-scarce environments. By combining quantum-inspired transformers with QVAE for synthetic data augmentation, Q-ASC achieved a significant boost in classification accuracy, reaching 68.3\% to 88.5\% on the TUT Acoustic Scenes 2016 dataset and outperforming the existing state-of-the-art by over 5\% in the best case. These results signified a substantial advancement in the field of acoustic scene classification, offering promising applications in industrial monitoring, environmental sound analysis, and healthcare. Future research should focus on further enhancing Q-ASC’s capabilities through the integration of self-supervised learning techniques and multi-modal data fusion. Such advancements could improve the classifier’s robustness and adaptability across diverse and complex acoustic environments, thereby extending its applicability and effectiveness in real-world scenarios.

\balance
\bibliography{abbriviation, Reference}
\bibliographystyle{IEEEtran}

\end{document}